\documentclass[12pt,a4paper]{article}
\usepackage{latexsym,amssymb,amsmath}
\usepackage[dvips]{graphicx}
\usepackage{calc}
\usepackage{times}
\newcommand{\bequ}{\begin{equation}}
\newcommand{\eequ}[1]{\label{#1}\end{equation}}
\newcommand\eq[1] {(\ref{#1})}
\DeclareMathOperator*{\ceil}{ceil}
\DeclareMathOperator*{\sign}{sign}
\DeclareMathOperator*{\erf}{erf}
\newcommand{\xavr}[1]{<\!\! #1\!\!>}
\newcommand{\xtavr}[1]{<\!\!<\!\! #1\!\!>\!\!>}
\newcommand{\solidspace}[1]{$\left.\right.$\hspace{#1}}

\newcounter{hours}\newcounter{minutes}

\title{Detailed analysis of a pseudoresonant interaction 
between cellular flames and velocity turbulence}
\author{V. Karlin}
\date{}

\begin{document}

\DeclareGraphicsExtensions{.eps}
\thispagestyle{empty}

\maketitle 

\begin{center}
Centre for Research in Fire and Explosions\\
University of Central Lancashire,
Preston PR1 2HE, UK\\
Email: VKarlin@uclan.ac.uk
\end{center}

\begin{abstract} 
This work is dedicated to the analysis of the delicate details 
of the effect of upstream velocity fluctuations on the flame 
propagation speed. The investigation was carried out using the 
Sivashinsky model of cellularisation of hydrodynamically unstable 
flame fronts. We identified the perturbations of the steadily 
propagating flames which can be significantly amplified over 
finite periods of time. These perturbations were used to model 
the effect of upstream velocity fluctuations on the flame front 
dynamics and to study a possibility to control the flame 
propagation speed.
\end{abstract}

\thispagestyle{empty}

\vspace{3mm}\noindent
{\bf Key words:} hydrodynamic flame instability, Sivashinsky 
equation, nonmodal amplification, flame-turbulence interaction

\vspace{3mm}\noindent
{\bf AMS subject classification:} 35S10, 76E17, 80A25, 65F15

\vspace{3mm}\noindent
{\bf Abbreviated title:} Analysis of pseudoresonant flame-turbulence 
interaction

\newpage

\section{Introduction}
Experiments show that cellularisation of flames results in
an increase of their propagation speed. In order to understand 
and exploit this phenomenon, we study the evolution of flame 
fronts governed by the Sivashinsky equation 
\bequ
\partial_{t}\Phi -2^{-1}\left(\partial_{x}\Phi\right)^{2} 
=\partial_{xx}\Phi
-(\gamma/2)\partial_{x}\mathcal{H}[\Phi]+f(x,t),\qquad
-\infty < x < \infty,\quad t>0
\eequ{SivaEq}
with the force term $f(x,t)$. Here $\Phi(x,t)$ is the 
perturbation of the plane flame front,
$\mathcal{H}[\Phi]=\pi^{-1}\int_{-\infty}^{\infty}(x-\xi)^{-1}\Phi(\xi,t)d\xi$
is the Hilbert transformation, and $\gamma=1-\rho_{b}/\rho_{u}$ 
is the contrast in densities of burnt and unburnt gases
$\rho_{b}$ and $\rho_{u}$ respectively. Initial perturbation
$\Phi(x,0)$ is given.

The equation without the force term was obtained in 
\cite{Sivashinsky77a} as an asymptotic mathematical model of 
cellularisation of flames subject to the hydrodynamic flame 
instability. The force term was suggested in \cite{Joulin88} 
in order to account for the effect of the upstream turbulence
on the flame front. It is equal to the properly scaled turbulent
fluctuations of the velocity field of the unburned gas.
In \cite{Joulin-Cambray92a} and \cite{Cambray-Joulin92} equation
\eq{SivaEq} was further refined in order to include effects
of the second order in $\gamma$. However, as mentioned in 
\cite{Cambray-Joulin92}, this modification can be compensated 
upon a Galilean transformation combined with a nonsingular
scaling. Thus, we have chosen to remain within the first order 
of accuracy in $\gamma$ of the original Sivashinsky model 
\eq{SivaEq} as it should have the same qualitative properties 
as the more quantitatively accurate one.

The asymptotically stable solutions to the Sivashinsky equation 
with $f(x,t)\equiv 0$ corresponding to the steadily propagating 
cellular flames do exist and are given by formula 
\bequ
\Phi_{N,L}(x,t)=V_{N,L}t 
+2\sum\limits_{n=1}^{N}\ln|\cosh 2\pi b_{n}/L-\cos 2\pi x/L|, 
\eequ{PoleSol} 
discovered in \cite{Thual-Frisch-Henon85}. Here, real $L>0$ 
and integer $N$ from within the range 
$0\le N\le N_{L}=\ceil(\gamma L/8\pi+1/2)-1$ are otherwise 
arbitrary parameters. Also, 
$V_{L}=2\pi NL^{-1}\left(\gamma-4\pi NL^{-1}\right)$, and
$b_{1},b_{2},\ldots,b_{N}$ satisfy a system of nonlinear 
algebraic equations available elsewhere. Functions \eq{PoleSol} 
have a distinctive set of $N$ complex conjugate pairs of poles 
$z_{n}=\pm ib_{n}$, $n=1,\ldots,N$ and are called the steady 
coalescent pole solutions respectively.

The steady coalescent pole solutions \eq{PoleSol} with the 
maximum possible number $N=N_{L}$ of the poles were found 
to be asymptotically, for $t\rightarrow\infty$, stable if 
the wavelength of the perturbations does not exceed $L$, 
see \cite{Vaynblat-Matalon00a}. However, in spite of their 
asymptotic stability, there are perturbations of these 
solutions which can be hugely amplified over finite 
intervals of time resulting in significant transients, see 
\cite{Karlin02a}. These perturbations are nonmodal, because 
they cannot be represented by the single eigenmodes of the 
linearised Sivashinsky equation. In what follows we are 
interested in solutions \eq{PoleSol} with $N=N_{L}$ and 
retain the index $L$ only. Also, in all reported 
calculations $\gamma=0.8$.

In this work we calculate the most amplifiable nonmodal 
perturbations to the asymptotically stable cellular 
solutions of the Sivashinsky equation and use them to 
investigate the response of the flame front to forcing. 
In particular, we study the effect of stochastic forcing or 
noise. The investigation of the effect of noise in the 
Sivashinsky equation was carried out numerically and the 
observations were reinforced by the analytical analysis of 
an approximation to the linearised Sivashinsky equation 
suggested in \cite{Joulin89b}.

\section{The largest growing perturbations}\label{WorstPert}
Substituting $\Phi(x,t)=\Phi_{L}(x,t)+\phi(x,t)$ into \eq{SivaEq} 
for $f(x,t)\equiv 0$ and linearising it with respect to the
$L$-periodic perturbations $\phi(x,t)$, one obtains
\bequ
\left\{\begin{array}{l}
\partial_{t}\phi=(\partial_{x}\Phi_{L})\partial_{x}\phi
+\partial_{xx}\phi
-(\gamma/2)\partial_{x}{\cal H}[\phi]=A_{L}\phi,\\
 \\
\phi(x,0)=\Phi(x,0)-\Phi_{L}(x,0).
\end{array}\right.
\eequ{LinSivaEq}
The operator $A_{L}$ generates the evolution operator $e^{tA_{L}}$, 
which provides the solution to \eq{LinSivaEq} in the form  
$\phi(x,t)=e^{tA_{L}}\phi(x,0)$. 

Assuming that the polar decomposition of the evolution 
operator does exist, we write it as 
\bequ
e^{tA_{L}}=\mathcal{U}(t)\mathcal{S}(t),
\eequ{polexp}
where $\mathcal{U}(t)$ is a partially isometric and $\mathcal{S}(t)
=\left[\left(e^{tA_{L}}\right)^{*}e^{tA_{L}}\right]^{1/2}$ 
is the nonnegative self-adjoint operator, see e.g. 
\cite{Gohberg-Krein}. The partial isometry of $\mathcal{U}(t)$ 
implies that it preserves the norm when mapping between the 
sets of values of $\left(e^{tA_{L}}\right)^{*}$ and $e^{tA_{L}}$, 
i.e. $\|\mathcal{U}(t)\phi\|=\|\phi\|$. Then, under certain 
conditions, $\|\phi(x,t)\|=\|\mathcal{S}(t)\phi(x,0)\|$ and for 
the $2$-norm the 
$\sup\limits_{\phi(x,0)\in\mathcal{D}\left(e^{tA_{L}}\right)}
\left\{\|\mathcal{S}(t)\phi(x,0)\|\right.$ 
$\left.\times\|\phi(x,0)\|^{-1}\right\}$ is equal to the 
largest eigenvalue $\sigma_{1}(t)$ of $\mathcal{S}(t)$. This
eigenvalue is associated with the eigenvector
$\psi_{1}(x,t)$ of $\mathcal{S}(t)$. 

The eigenvectors $\psi_{\alpha}(x,t)$ of $\mathcal{S}(t)$ are 
mutually orthogonal at any given time $t=t^{*}$ and can be used 
as a basis in the space of the admissible initial conditions
$\phi(x,0)$ $=\sum\limits_{\alpha=1}^{\infty}c_{\alpha}(0,t^{*})$
$\times\psi_{\alpha}(x,t^{*})$. Then, the associated eigenvalues 
$\sigma_{\alpha}(t^{*})$ provide the magnitudes of
amplification of the $\psi_{\alpha}(x,t^{*})$ components of
the initial condition $\phi(x,0)$ by the time instance $t^{*}$.
Note, that for \eq{LinSivaEq} the $2$-norm of the perturbation 
$\phi(x,t)$ is just its energy and that the eigenvalues 
$\sigma_{\alpha}(t)$, $\alpha=1,2,\ldots$ and eigenvectors 
$\psi_{\alpha}(x,t)$ of $\mathcal{S}(t)$ are the singular 
values and the right singular vectors of $e^{tA_{L}}$ 
respectively.

According to \cite{Karlin04a}, the Fourier image
$\widetilde{A_{L}}$ of the operator $A_{L}$
is defined by the $(k,l)$-th entry of its double infinite
($-\infty<k,l<\infty$) matrix
\bequ
(\widetilde{A_{L}})_{k,l}
=\left(-\frac{4\pi^{2}}{L^{2}}k^{2}
+\frac{\pi\gamma}{L}|k|\right)\delta_{k,l}
+\frac{8\pi^{2}}{L^{2}}l\sign(k-l)
\sum\limits_{n=1}^{N_{L}}e^{-2\pi b_{n}|k-l|/L},
\eequ{2_2e}
where $\delta_{k,l}$ is the Kronecker's symbol. By limiting
our consideration to the first $K$ harmonics, we approximate
our double infinite matrix $\widetilde{A_{L}}$ with
the $(2K+1)\times(2K+1)$ matrix $\widetilde{A_{L}^{(K)}}$,
whose entries coincide with those of $\widetilde{A_{L}}$
for $-K\le k,l\le K$. Then, the matrix
$e^{t\widetilde{A_{L}^{(K)}}}\approx\widetilde{e^{tA_{L}}}$
can be effectively evaluated by the scaling and squaring
algorithm with a Pad\'{e} approximation.
Eventually, the required estimations of $\sigma_{\alpha}(t)$
and Fourier images of $\psi_{\alpha}(x,t)$ can be obtained
through the singular value decomposition (SVD) of
$e^{t\widetilde{A_{L}^{(K)}}}$, see e.g. \cite{Golub-VanLoan89}. 

Indeed, if the SVD of $e^{t\widetilde{A_{L}^{(K)}}}$ is 
given by
\bequ
e^{t\widetilde{A_{L}}^{(K)}}
=\mathcal{W}(t)\mathcal{D}(t)\mathcal{V}(t)^{*},
\eequ{svdexp}
where $\mathcal{W}(t)$, $\mathcal{V}(t)$ are unitary and 
$\mathcal{D}(t)$ is the nonnegative diagonal matrix, then 
the matrices 
\bequ
\mathcal{U}(t)=\mathcal{W}(t)\mathcal{V}(t)^{*},\qquad
\mathcal{S}(t)=\mathcal{V}(t)\mathcal{D}(t)\mathcal{V}(t)^{*}
\eequ{usexp}
satisfy the adequate finite-dimensional projection of the 
polar decomposition \eq{polexp} and the eigenvalues 
$\sigma_{\alpha}(t)$, $\alpha=1,2,\ldots$ and eigenvectors 
$\psi_{\alpha}(x,t)$ of $\mathcal{S}(t)$ are just the singular 
values and the Fourier syntheses of the right singular vectors 
of $e^{t\widetilde{\mathcal{A}_{L}}^{(K)}}$ respectively. 

\begin{figure}[ht]\begin{centering}
  \includegraphics[height=58.5mm,width=90mm]{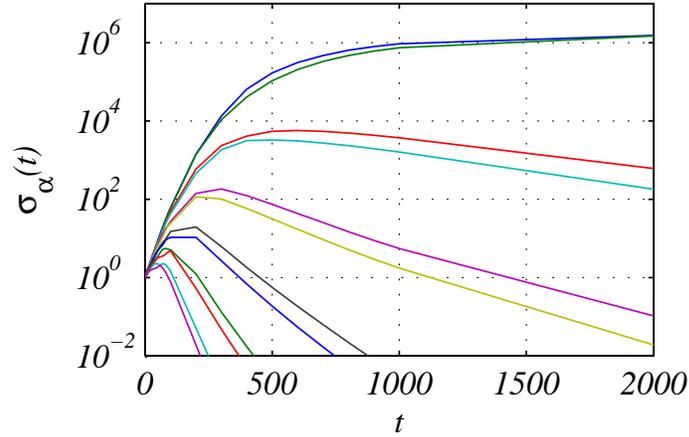}
  \caption{Twelve largest singular values of $e^{tA_{L}}$.}
  \label{sgm}
\end{centering}\end{figure}

Graphs showing dependence of a few largest singular values
of $e^{tA_{L}}$ versus time are shown in Fig. \ref{sgm}.
One may see that values of
$\sigma_{1,2}(t)$ for large enough $t$ match the estimation
of the largest possible amplification of the perturbations
$\phi(x,t)$ obtained in \cite{Karlin02a} by a different method.
An even more impressive observation is that the dimension of
the subspace of the significantly amplifiable perturbations is
very low. Perturbations of only two types can be amplified by
about $10^{6}$ times.
 
\begin{figure}[ht]\begin{centering}
  \includegraphics[height=144mm,width=120mm]{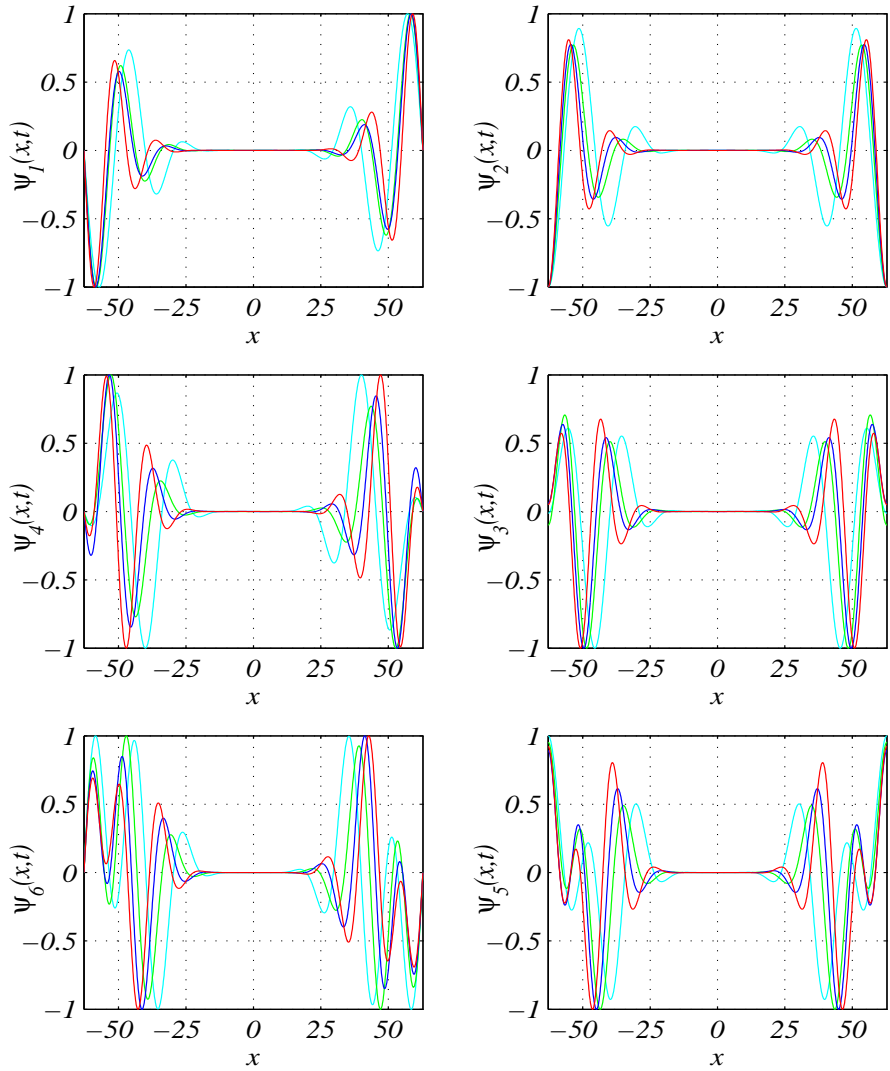}
  \caption{Right singular vectors of $e^{tA_{L}}$ corresponding to 
           the six largest $\sigma_{\alpha}(t)$ for $t=100$ (cyan),
           $200$ (green), $300$ (blue), and $10^{3}$ (red).}
  \label{psi}
\end{centering}\end{figure}

The initial conditions $\phi(x,0)$, which would be the most
amplified once by $t^{*}=100$, $200$, $300$ and $10^{3}$, i.e. 
$\psi_{\alpha}(x,t^{*})$, are depicted in Fig. \ref{psi}.
The dominating singular modes $\psi_{\alpha}(x,t)$ stabilize 
to some limiting functions for $t>300$. For example, their 
graphs for $t=500$ and $t=10^{3}$ are indistinguishable in 
Fig. \ref{psi}. However, they vary in time
significantly when $t<300$ and for $t=200$ the associated
amplification $\sigma_{1,2}(200)$ is already about $10^{3}$,
though $\psi_{1,2}(x,200)$ does not coincide with neither
$\psi_{1,2}(x,10^{3})$ nor $\psi_{3,4}(x,10^{3})$. Thus, the
dependence of $\psi_{\alpha}$ on time makes the dimension
of the subspace of perturbations, which can be amplified say
about $10^{3}$ times much higher than two in contrast to
what could be concluded from the graphs in Fig. \ref{sgm}. 
This illustrates the complicatedness of studies of the effect 
of transient amplification on short time scales $t<300$.

\begin{figure}[ht] 
\centerline{\includegraphics[height=50mm,width=80mm]{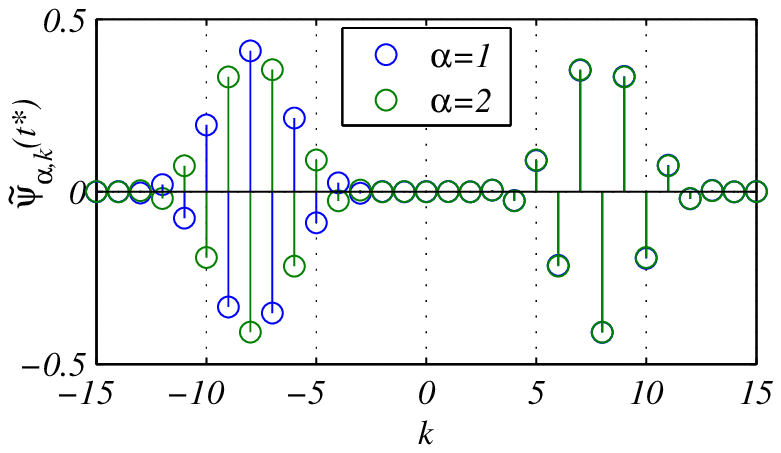}}
\caption{Fourier coefficients of $\psi_{\alpha}(x,t^{*}) 
=\sum\limits_{k=-\infty}^{\infty}\widetilde{\psi}_{\alpha,k}(t^{*})
e^{i2\pi kx/L}$ for $t^{*}=10^{3}$.}\label{psiFour}
\end{figure}

Fourier components of $\psi_{1}(x,t^{*})$ and $\psi_{2}(x,t^{*})$ 
for $t^{*}=10^{3}$ are depicted in Fig. \ref{psiFour}. 

\begin{figure}[ht]
\begin{center}
\begin{tabular}{lll}
  \includegraphics[height=40mm,width=60mm]{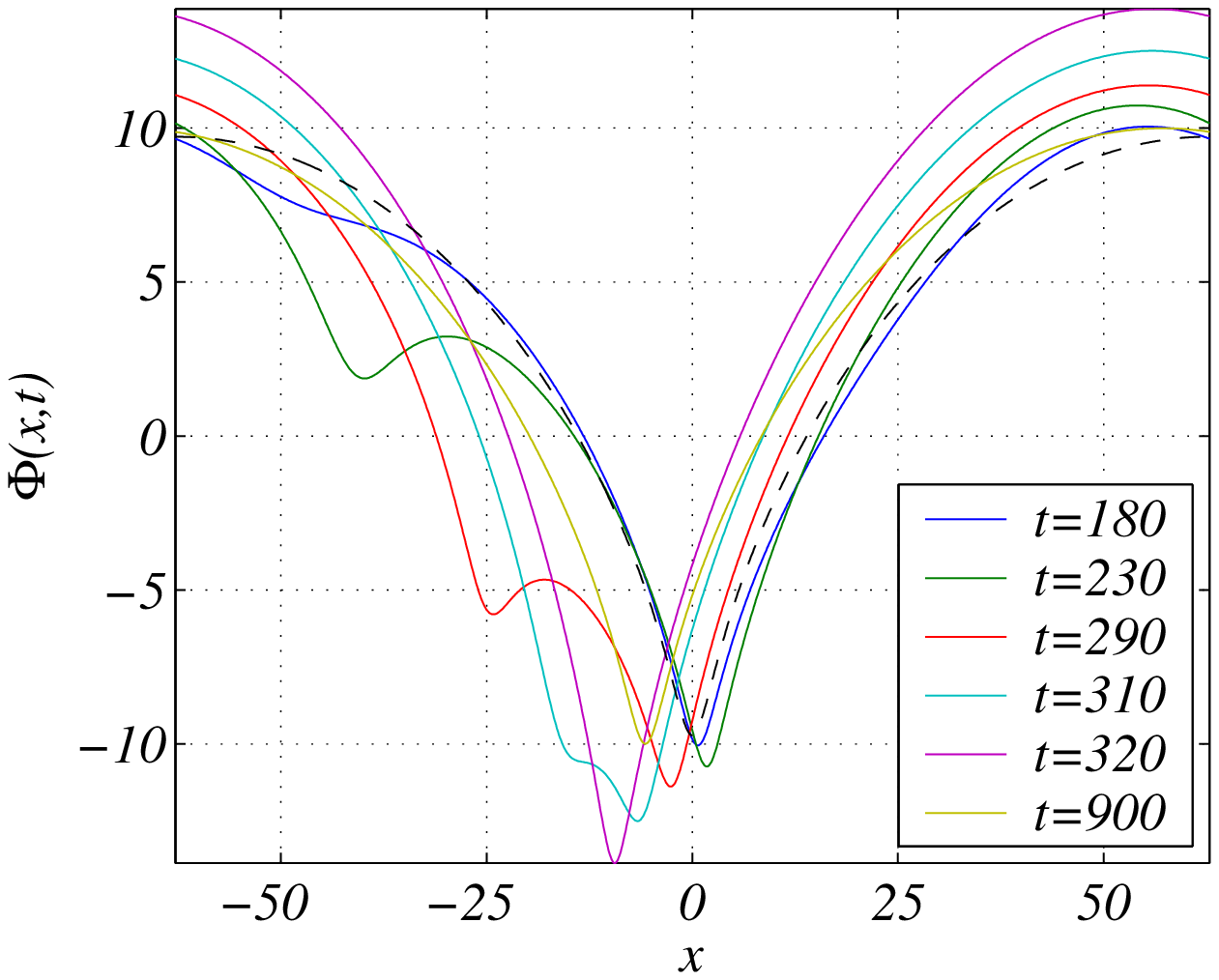} &
  \solidspace{0mm} &
  \includegraphics[height=40mm,width=60mm]{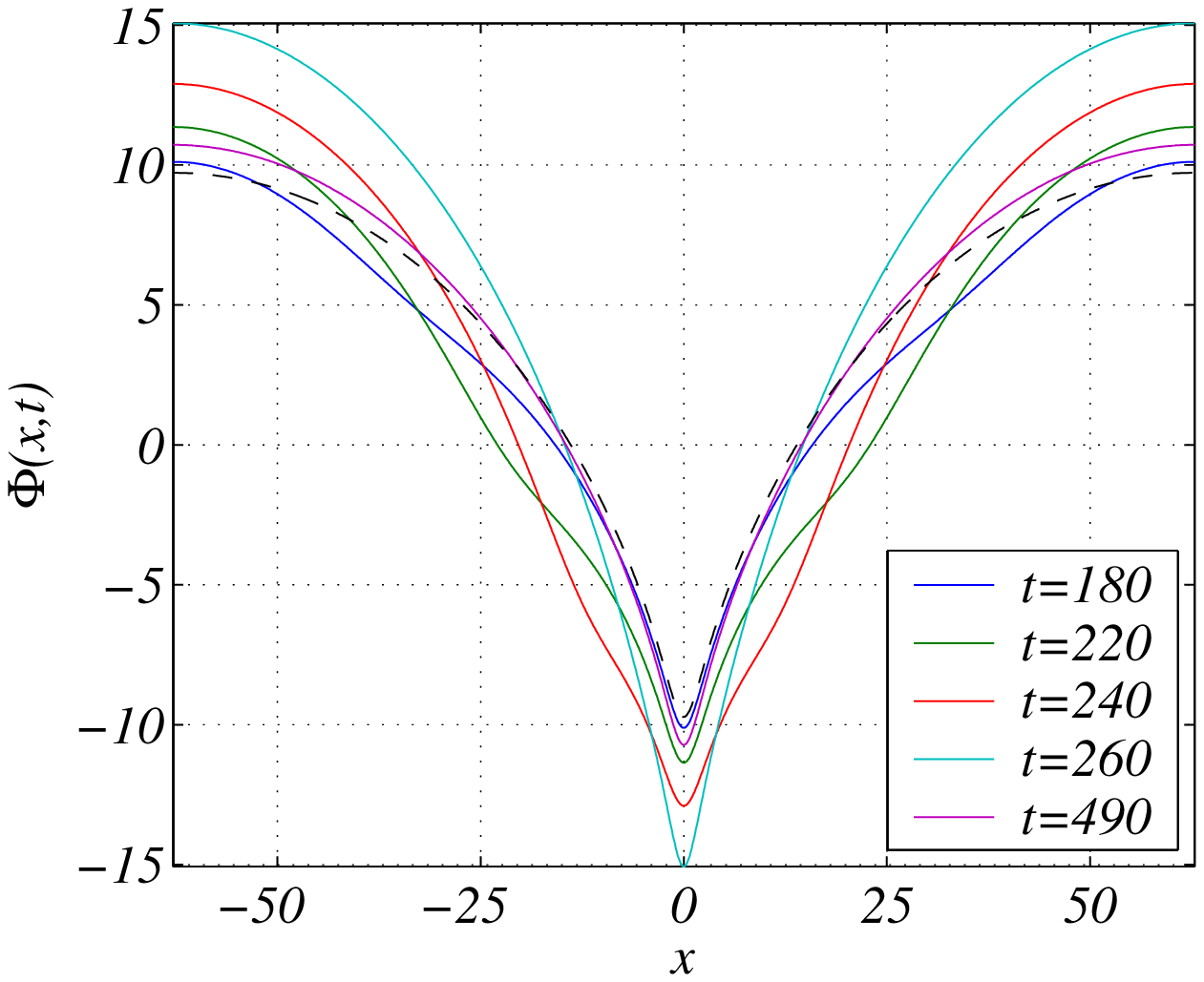}\\
  \includegraphics[height=16mm,width=60mm]{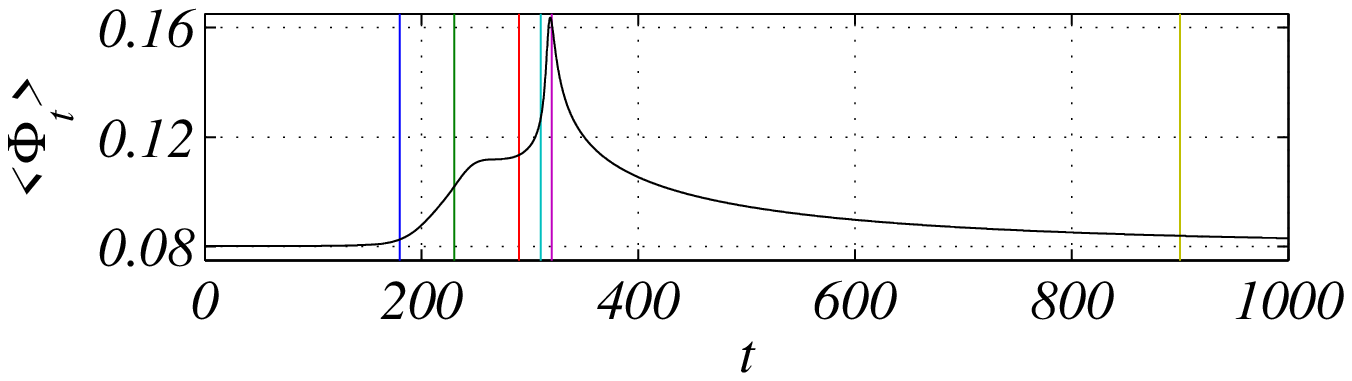} &
  \solidspace{0mm} &
  \includegraphics[height=16mm,width=60mm]{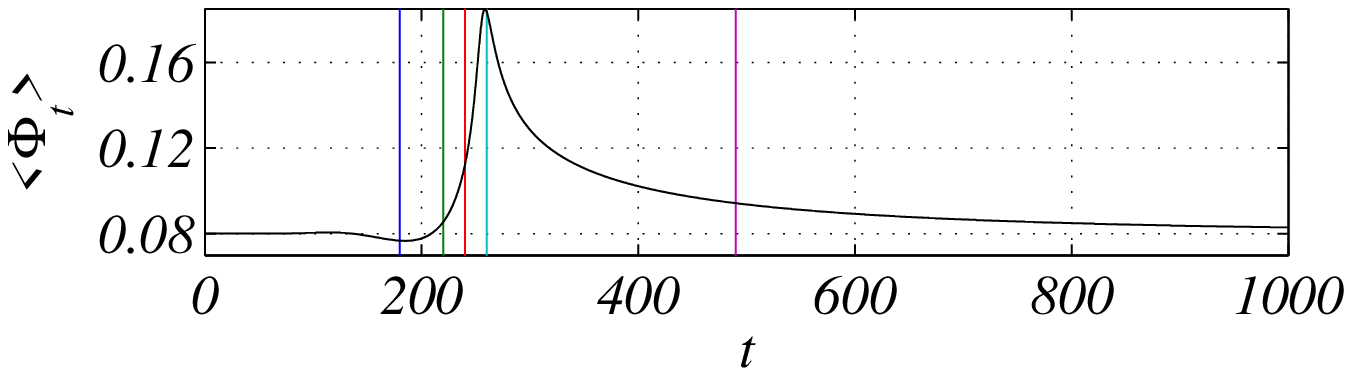}\\
  \solidspace{5mm}(a) $\varepsilon=10^{-3}$, $\alpha=1$ & & 
  \solidspace{5mm}(b) $\varepsilon=-10^{-3}$, $\alpha=2$\\
   & & \\
  \includegraphics[height=40mm,width=60mm]{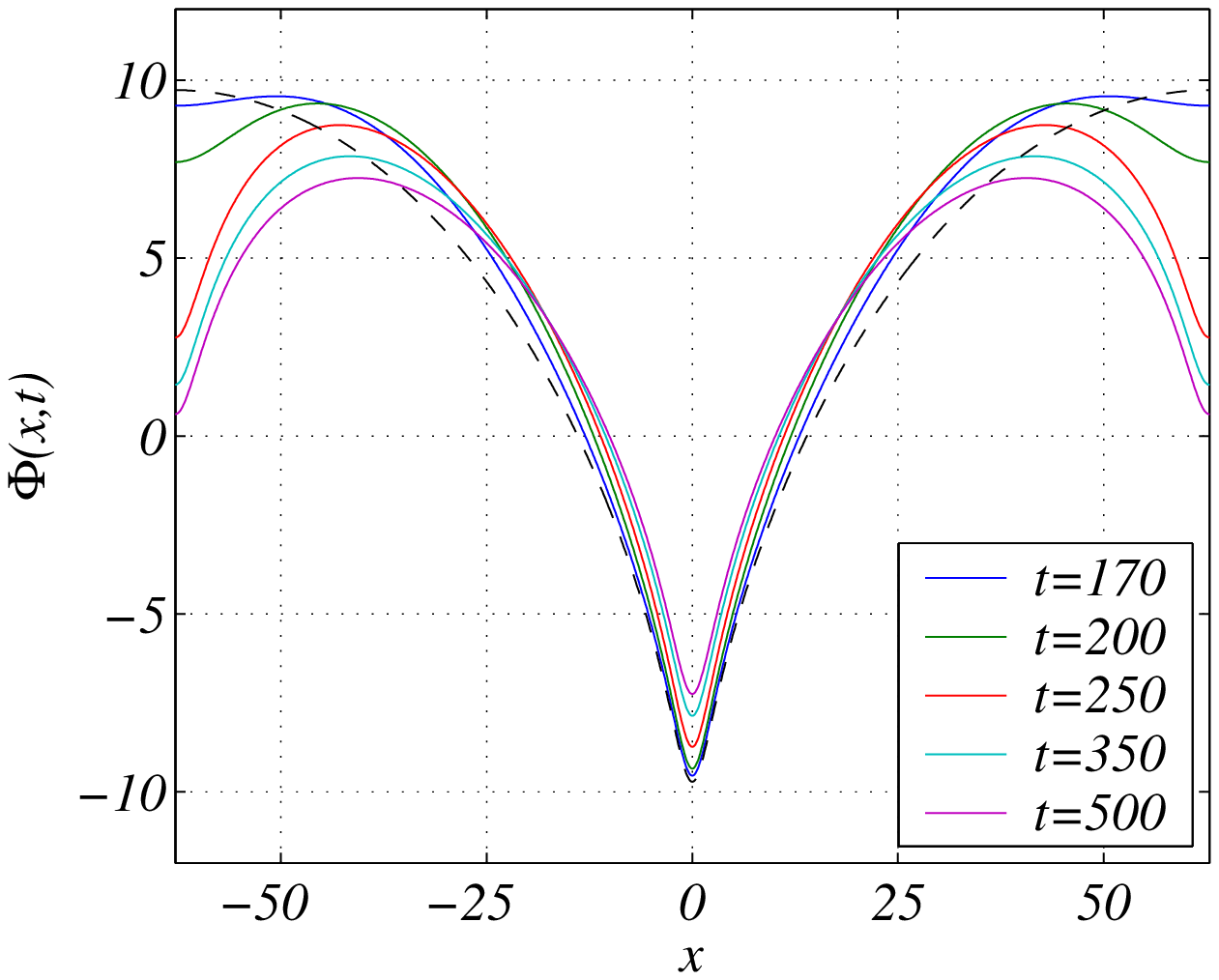} &
  \solidspace{0mm} &
  \includegraphics[height=40mm,width=60mm]{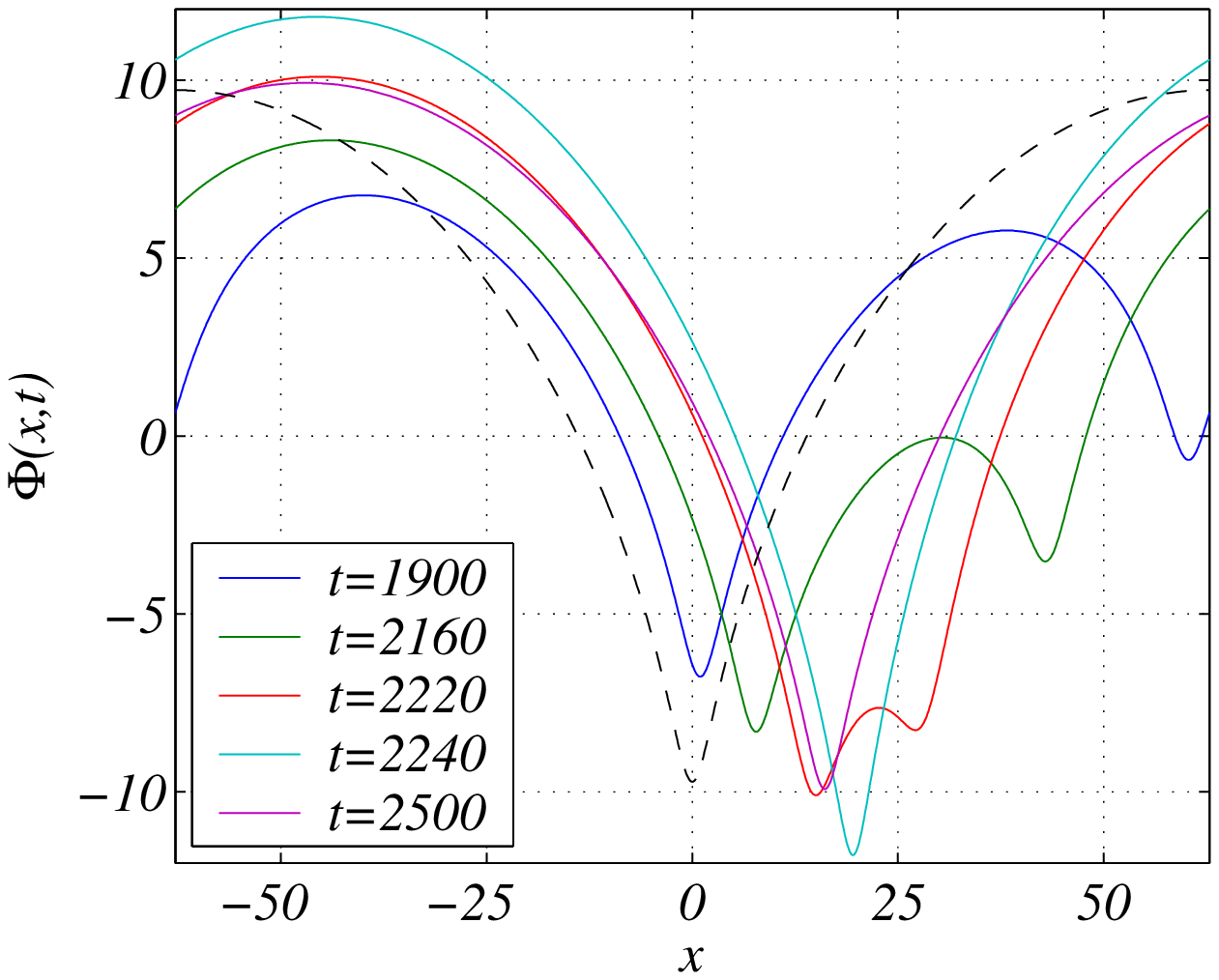}\\
  \solidspace{1.5mm}\includegraphics[height=16mm,width=60mm]{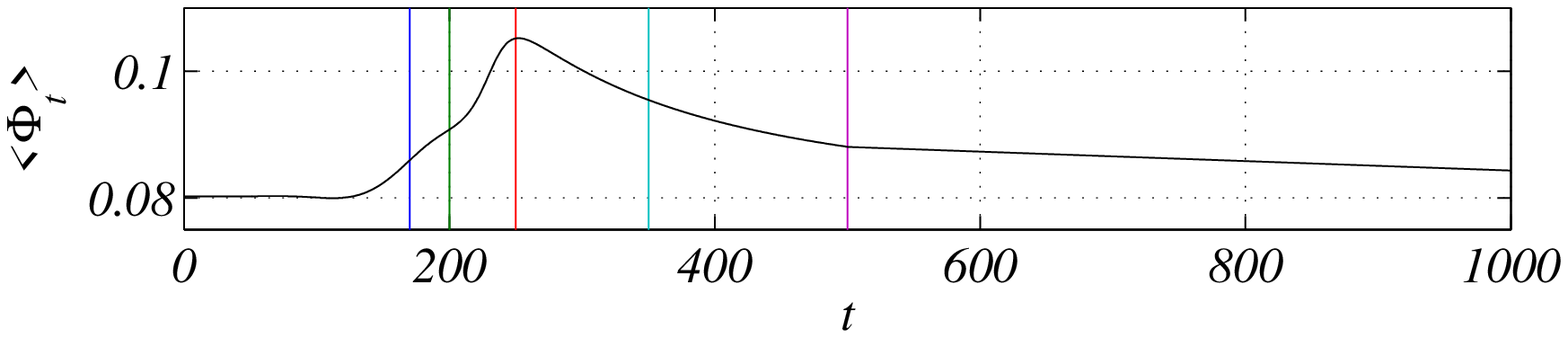} &
  \solidspace{0mm} &
  \solidspace{1.5mm}\includegraphics[height=16mm,width=60mm]{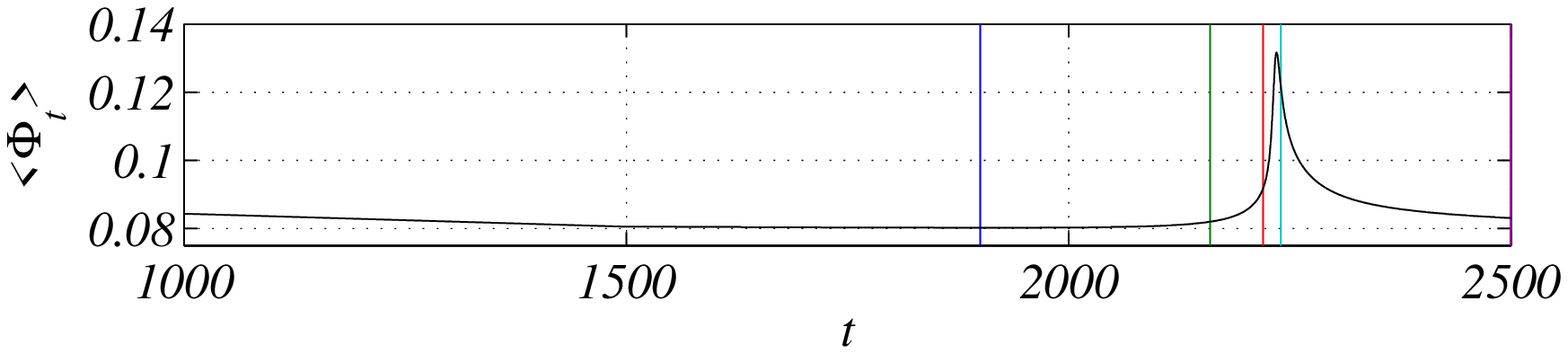}\\
  \solidspace{5mm}(c) $\varepsilon=10^{-3}$, $\alpha=2$ & & 
  \solidspace{5mm}(d) $\varepsilon=10^{-3}$, $\alpha=2$, continues (c)\\
\end{tabular}
\end{center}
\caption{Solutions to \eq{SivaEq} for $\Phi(x,0)=\Phi_{L}
         +\varepsilon\psi_{\alpha}(x,t^{*})$ and $L=40\pi$.}
         \label{mode1p}
\end{figure}

Evolution of the perturbations, which grow the most and is  
governed by the nonlinear Sivashinsky equation, is illustrated 
in Fig. \ref{mode1p}. All the profiles were displaced 
vertically in order to compensate for steady propagation of 
flames in such a way that their spatial averages are equal to zero. 
Matching graphs of the spatially averaged flame propagation speed 
\bequ 
<\Phi_{t}> = \frac{1}{L}\int\limits_{-L/2}^{L/2}\partial_{t}\Phi(x,t)dx, 
\eequ{FlameSpeed} 
are shown as well. The initial conditions were
$\Phi(x,0)=\Phi_{L}(x,0)+\varepsilon\psi_{\alpha}(x,t^{*})$, where 
$\varepsilon=\pm 10^{-3}$, $\alpha=1,2$, and $t^{*}=10^{3}$. The
computational method used in this work was presented in
\cite{Karlin-Maz'ya-Schmidt03}.

The asymmetric singular mode $\psi_{1}(x,t^{*})$ results in
appearance of a small cusp to the left or to the right from
the trough of $\Phi_{L}(x,0)$ depending on the sign of
$\varepsilon$. After the cusp merges with the trough, the 
flame profile converges slowly to $\Phi_{L}(x+\Delta x,t)$,
where $\sign(\varepsilon)\Delta x>0$. For a positive 
$\varepsilon=10^{-3}$ the effect is illustrated in 
Fig. \ref{mode1p}. Graphs of $\Phi(x,t)$ for $\varepsilon=-10^{-3}$
are exact mirror reflections of those depicted in \ref{mode1p}(a)
and graphs of $<\Phi_{t}>$ are exactly the same.  

The symmetric singular mode $\psi_{2}(x,t^{*})$ produces two 
symmetric dents moving towards the trough on both sides of
the profile if $\varepsilon<0$, see Fig. \ref{mode1p}(b). By
$t\approx500$ the flame profile returns very closely to 
$\Phi_{L}(x,t)$. For $\varepsilon>0$ two small cusps move 
towards the boundaries of the computational domain creating a 
quasi-steady structure shown in Fig. \ref{mode1p}(c) for 
$t=270$. This structure survives until $t\approx1800$, but 
eventually bifurcates, see Fig. \ref{mode1p}(d), and the 
solution converges to $\Phi_{L}(x+\Delta x,t)$, $\Delta x<0$. 
It looks like the bifurcation in question is associated 
with the lack of the asymptotic stability of the intermediate 
quasi-steady structure. As such, it was triggered by a random 
perturbation and could equally result in the displacement of 
the limiting flame front profile into the opposite direction 
$\Delta x>0$.

Behavior of perturbations $\psi_{1,2}(x,t^{*})$ of the amplitude 
$\varepsilon=10^{-6}$ was not as impressive, but they managed to 
produce a visible effect on the flame front profile. The same can be said about 
$\psi_{3,4}(x,t^{*})$ of the amplitude $\varepsilon=10^{-3}$. 
Perturbations corresponding to $\psi_{\alpha}(x,t^{*})$ of higher orders
did not grow significantly and did not cause any noticeable changes 
to $\Phi_{L}$ for $\varepsilon$ up to $10^{-2}$.

Thus, the singular modes $\psi_{1,2}(x,t^{*})$ should be responsible for 
the interaction of the flame front $\Phi_{L}(x,t)$ with all the perturbations 
of small enough amplitude. The time scale of these interactions is about 
$300$ for $L=40\pi$ and is of order $O(L)$ in general. More 
singular modes $\psi_{\alpha}(x,t^{*})$ of higher orders $\alpha>2$ are 
becoming important as the amplitude of the perturbations grows. The time 
scale of evolution of $\phi(x,t)$ for $\phi(x,0)=\psi_{\alpha}(x,t^{*})$ 
lessens as $\alpha$ grows necessitating to take into account the dependence of 
$\psi_{\alpha}(x,t^{*})$ on $t^{*}$ and creating further problems in the 
efficient description of the subspace of important perturbations. Therefore, 
there is a critical perturbation amplitude beyond which the representation 
of $f(x,t)$ in terms of the singular modes $\psi_{\alpha}(x,t^{*})$ is 
not as beneficial as for smaller amplitudes.

\section{A simplified linear model}
Prior to experimenting with \eq{SivaEq} we consider a simplified 
linear model suggested in \cite{Joulin89b}. The $L$-periodic steady 
coalescent $N_{L}$-pole solution \eq{PoleSol} has a characteristic
wavy or cellular structure and can be represented in a vicinity of the 
crest as $\Phi_{L}(x,t)\approx \Phi_{L}(0,t)-x^{2}/(2R)+O(x^{4})$. 
Here, $R$ is the 
radius of curvature of the flame front profile in the crest. 
For large enough $L$, it can be approximated as $R=c_{1}L+c_{2}$, 
where $c_{1}$ and $c_{2}$ are some constants. 
Note, that in the approximation of $\Phi_{L}(x,t)$ the origin $x=0$ 
was chosen exactly in the crest of 
$\Phi_{L}(x,t)$. Thus, $\partial_{x}\Phi_{L}\approx -x/R+O(x^{3})$ 
in a vicinity of $x=0$ and the equation suggested in 
\cite{Joulin89b} can be written as 
\bequ
\partial_{t}\phi+R^{-1}x\partial_{x}\phi=\partial_{xx}\phi 
+(\gamma/2)\partial_{x}\mathcal{H}[\phi]+f(x,t),
\eequ{Joulin}
where $-\infty<x<\infty$ and $t>0$. The latter equation is much 
simpler than \eq{LinSivaEq}, yet it is meaningful enough to study 
the development of perturbations of $\Phi_{L}(x,t)$ appearing 
in the crest. 

Equation \eq{Joulin} can be solved exactly. Applying the 
Fourier transformation we obtain
\bequ
\partial_{t}\mathcal{F}[\phi]
-R^{-1}\xi\partial_{\xi}\mathcal{F}[\phi]
=-\left(4\pi^{2}\xi^{2}-\pi\gamma|\xi|
-R^{-1}\right)\mathcal{F}[\phi]+\mathcal{F}[f](\xi,t),
\eequ{ftJoulin}
which is a linear non-homogeneous hyperbolic equation of 
the first order. Using the standard method of characteristics
its exact solution can be written as follows 
\bequ
\mathcal{F}[\phi](\xi,t)
=\mathcal{G}(\xi,t)\mathcal{F}[\phi^{(0)}](|\xi|e^{t/R}) 
+\int\limits_{0}^{t}\mathcal{G}(\xi,t-\tau) 
\mathcal{F}[f]\left[|\xi|e^{(t-\tau)/R},\tau\right]d\tau, 
\eequ{convsol} 
where 
\bequ 
\mathcal{G}(\xi,t) 
=e^{t/R-2\pi^{2}R\left(e^{2t/R}-1\right)\xi^{2} 
+\pi\gamma R\left(e^{t/R}-1\right)|\xi|}, 
\eequ{funG1}
and $\mathcal{F}[f](\xi,t)=\int_{-\infty}^{\infty}
f(x,t)e^{-i2\pi x\xi}dx$ 
denotes the Fourier transformation of $f(x,t)$. 

\begin{figure}[ht]
\begin{center}
\begin{tabular}{lll}
  \includegraphics[height=40mm,width=60mm]{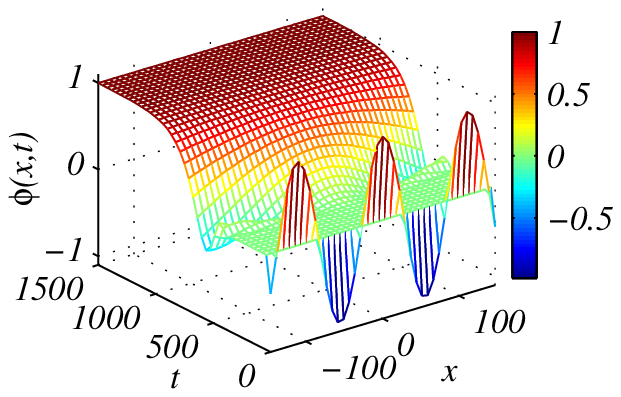} &
  \solidspace{5mm} &
  \includegraphics[height=40mm,width=60mm]{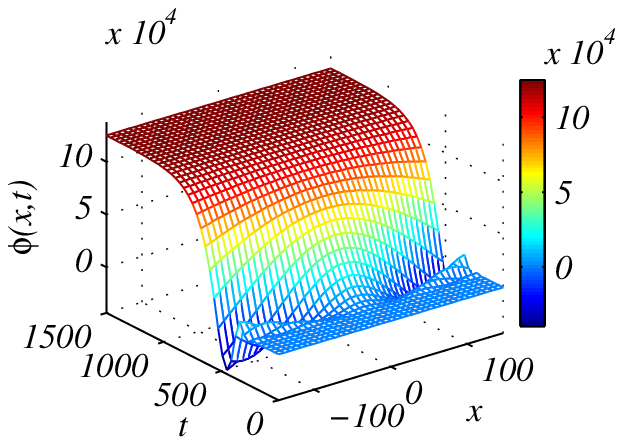}\\
  \solidspace{5mm}(a) $\xi_{0}=\gamma/(2\pi)$, $\varphi=0$ & & 
  \solidspace{5mm}(b) $\xi_{0}=\gamma/(4\pi)$, $\varphi=0$\\
   & & \\
  \includegraphics[height=40mm,width=60mm]{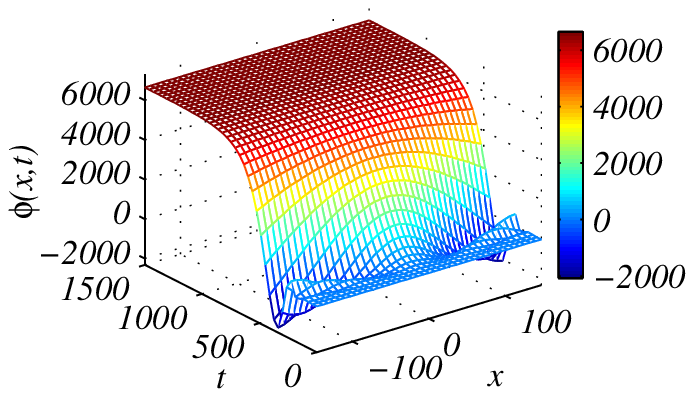} &
  \solidspace{5mm} &
  \includegraphics[height=40mm,width=60mm]{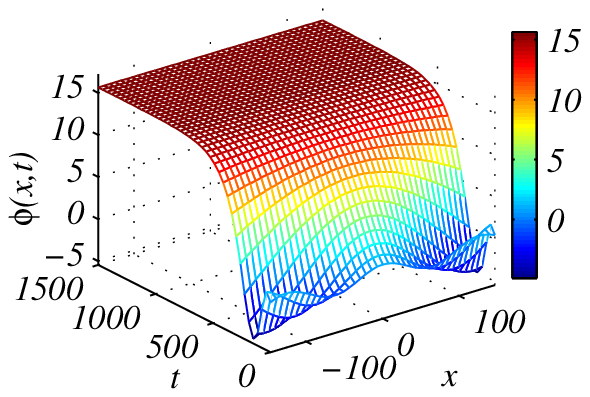}\\
  \solidspace{5mm}(c) $\xi_{0}=\gamma/(8\pi)$, $\varphi=0$ & & 
  \solidspace{5mm}(d) $\xi_{0}=\gamma/(32\pi)$, $\varphi=0$\\
   & & \\
  \includegraphics[height=40mm,width=60mm]{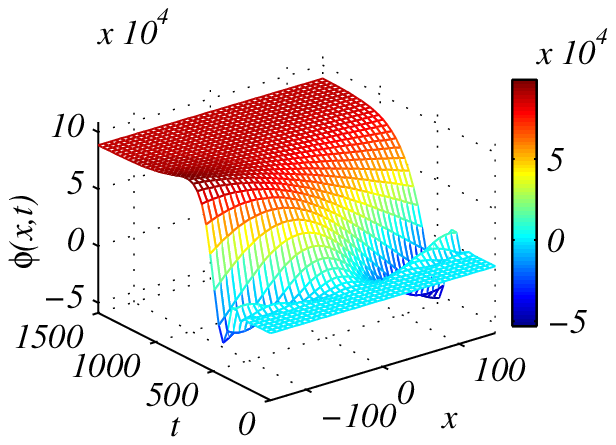} &
  \solidspace{5mm} &
  \includegraphics[height=40mm,width=60mm]{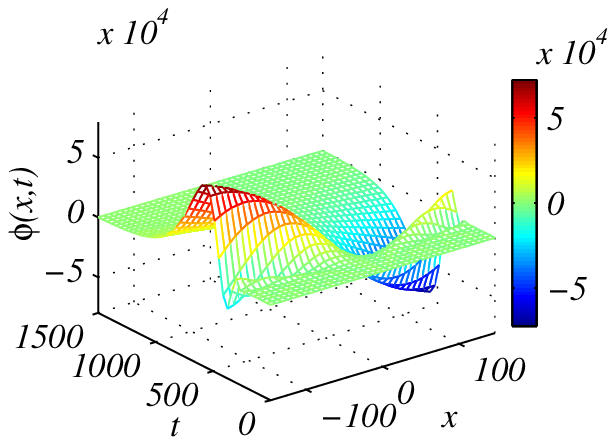}\\
  \solidspace{5mm}(e) $\xi_{0}=\gamma/(4\pi)$, $\varphi=\pi/4$ & & 
  \solidspace{5mm}(f) $\xi_{0}=\gamma/(4\pi)$, $\varphi=\pi/2$\\
\end{tabular}
\end{center}
\caption{Examples of solutions \eq{cosphisol} for $R=146.7126$, 
         which corresponds to $L=40\pi$.}
         \label{cossol4}
\end{figure}

If the initial condition is a single harmonics 
$\phi^{(0)}(x)=\cos(2\pi\xi_{0}x+\varphi)$ and $f(x,t)\equiv 0$, 
then 
\bequ
\phi(x,t)=e^{-2\pi^{2}R\left(1-e^{-2t/R}\right)\xi_{0}^{2} 
+\pi\gamma R\left(1-e^{-t/R}\right)\xi_{0}} 
\cos\left(2\pi\xi_{0}xe^{-t/R}+\varphi\right). 
\eequ{cosphisol} 
The infinite time limit of \eq{cosphisol} is equal 
$e^{-2\pi^{2}R\xi_{0}^{2}+\pi\gamma R\xi_{0}}\cos\varphi$ 
and is reached effectively on the time scale of order $O(R)$.
This time limit attains its maximum 
$e^{\gamma^{2}R/8}\cos\varphi$ for $\xi_{0}=\xi^{*}=\gamma/(4\pi)$, 
matching the asymptotic estimation of \cite{Joulin89b}.
Note that the wave number of the largest Fourier component
$k^{*}$ of both $\psi_{1}(x,t^{*})$ and $\psi_{2}(x,t^{*})$ for 
$t^{*}>300$ is equal to $\xi^{*}=k^{*}/L=\gamma/(4\pi)$ as well, 
see Fig. \ref{psiFour}. A few graphical examples of 
function \eq{cosphisol} are given in Fig. \ref{cossol4}. 
Note that the argument of the cosine in \eq{cosphisol} depends 
on time, which means that even if the initial condition 
$\phi^{(0)}(x)$ is a linear combination of mutually orthogonal 
cosine harmonics, then the solution $\phi(x,t)$ will remain a 
linear combination of cosine harmonics for $t>0$, but those 
harmonics will no longer be mutually orthogonal. This explains 
why the most amplified perturbations are formed by linear 
combinations of a few initially orthogonal harmonics and 
approximate $\psi_{1,2}(x,t^{*})$ asymptotically for 
$L\rightarrow\infty$.

Behaviour of \eq{cosphisol} is in a sharp contrast with 
the evolution of the single harmonics perturbations of the 
plane flame front
\bequ
\phi(x,t)=e^{\left(-4\pi^{2}\xi_{0}^{2}
+\pi\gamma\xi_{0}\right)t}\cos(2\pi\xi_{0}x+\varphi),
\eequ{planesol}
which grow infinitely if $\xi_{0}<\gamma/(4\pi)$ or decay 
otherwise. They are governed by the equation associated with 
a self-adjoint differential operator, which is obtained from 
\eq{Joulin} upon removal of the term $R^{-1}x\partial_{x}\phi$. 
Solution \eq{planesol} does not result from \eq{cosphisol} 
for $R\rightarrow\infty$, but is only 
equivalent to it when $t/R\ll 1$. The difference between
\eq{cosphisol} and \eq{planesol} is an explicit illustration
of the nonnormality of \eq{Joulin} introduced by the 
non-selfadjoint term $R^{-1}x\partial_{x}\phi$. Flattening 
of the crests of cellular flames and increasing local 
resemblance with the plane front as $R$ increases was 
noticed long time ago, prompting a hypothesis of a secondary 
Darrieus-Landau instability. Model \eq{Joulin} indicates that 
the hypothesis 
is unlikely to be correct. Although, because of the flattening 
of the crests of the flame front profile, perturbations of the 
front can be transiently amplified at a rate rapidly increasing 
with $R$, this transient amplification is entirely different 
from the infinite growth of perturbations in the Darrieus-Landau 
instability of plane flames. Moreover dynamics of 
perturbations in the case of cellular flames does not 
converge to that of the plane ones continuously in the 
limit $R\rightarrow\infty$.

\begin{figure}[ht]
\begin{center}
\begin{tabular}{lll}
  \includegraphics[height=40mm,width=60mm]{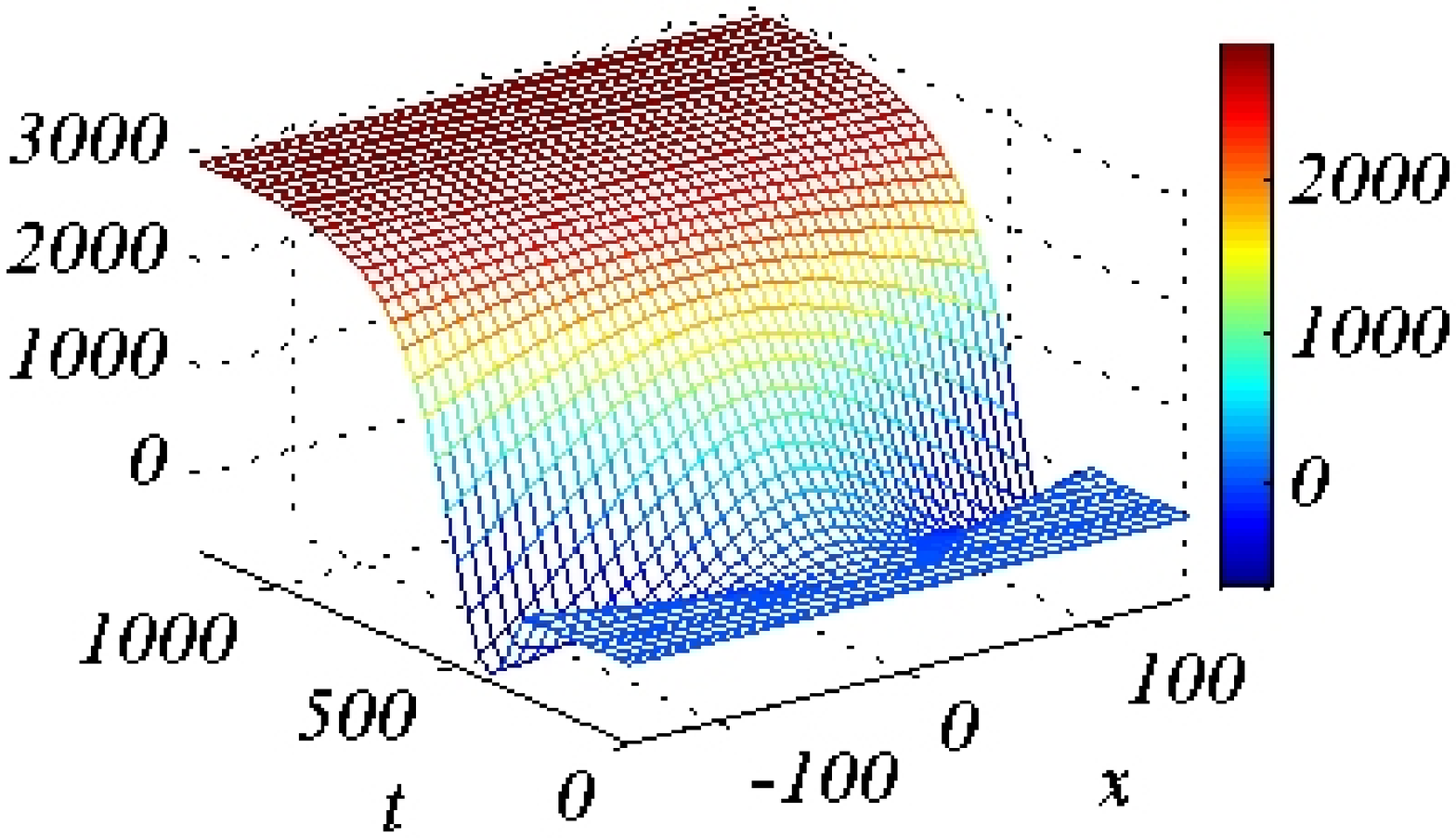} &
  \solidspace{5mm} &
  \includegraphics[height=40mm,width=60mm]{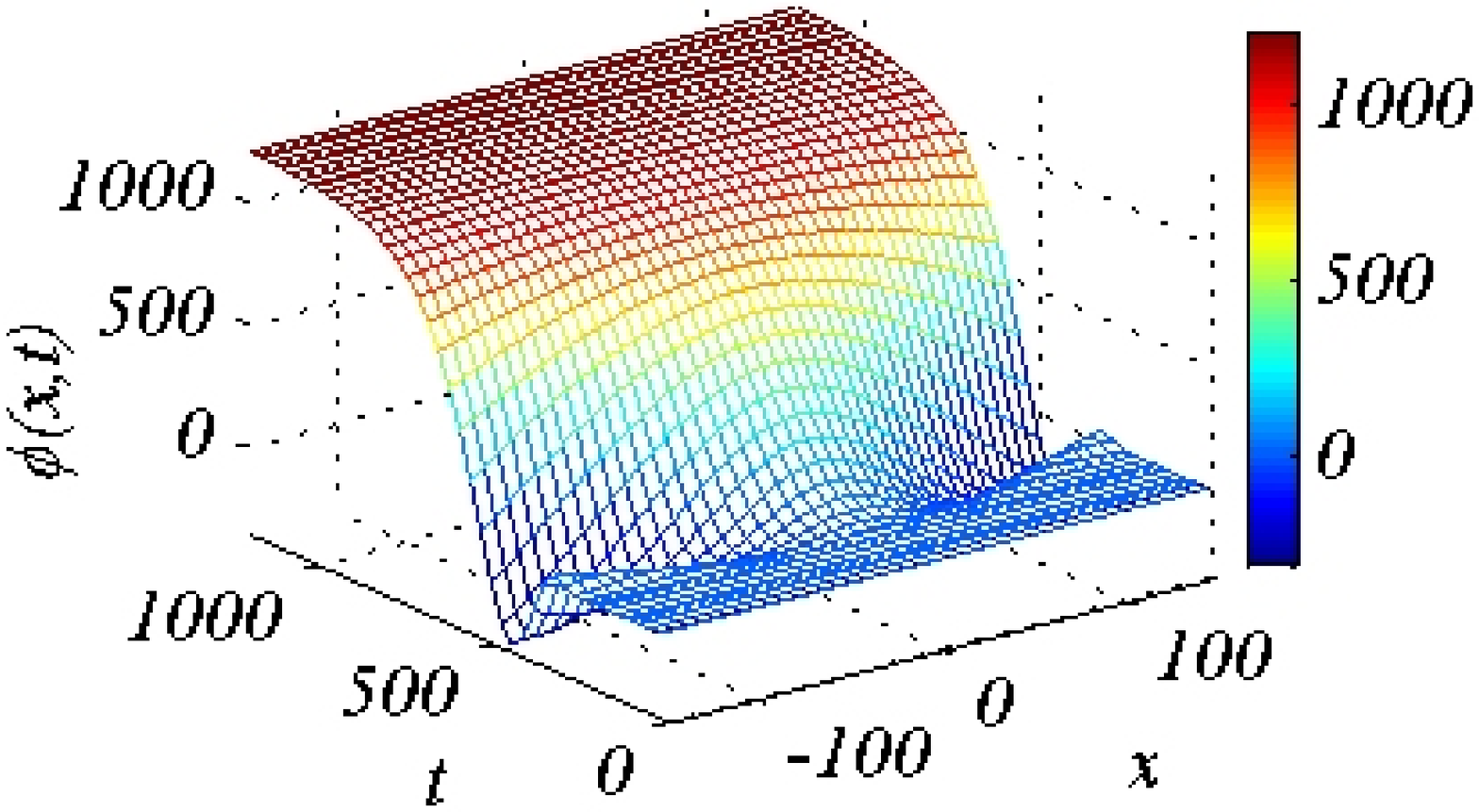}\\
  \solidspace{5mm}(a) $p=1/R$ & & 
  \solidspace{5mm}(b) $p=R$\\
   & & \\
  \includegraphics[height=40mm,width=60mm]{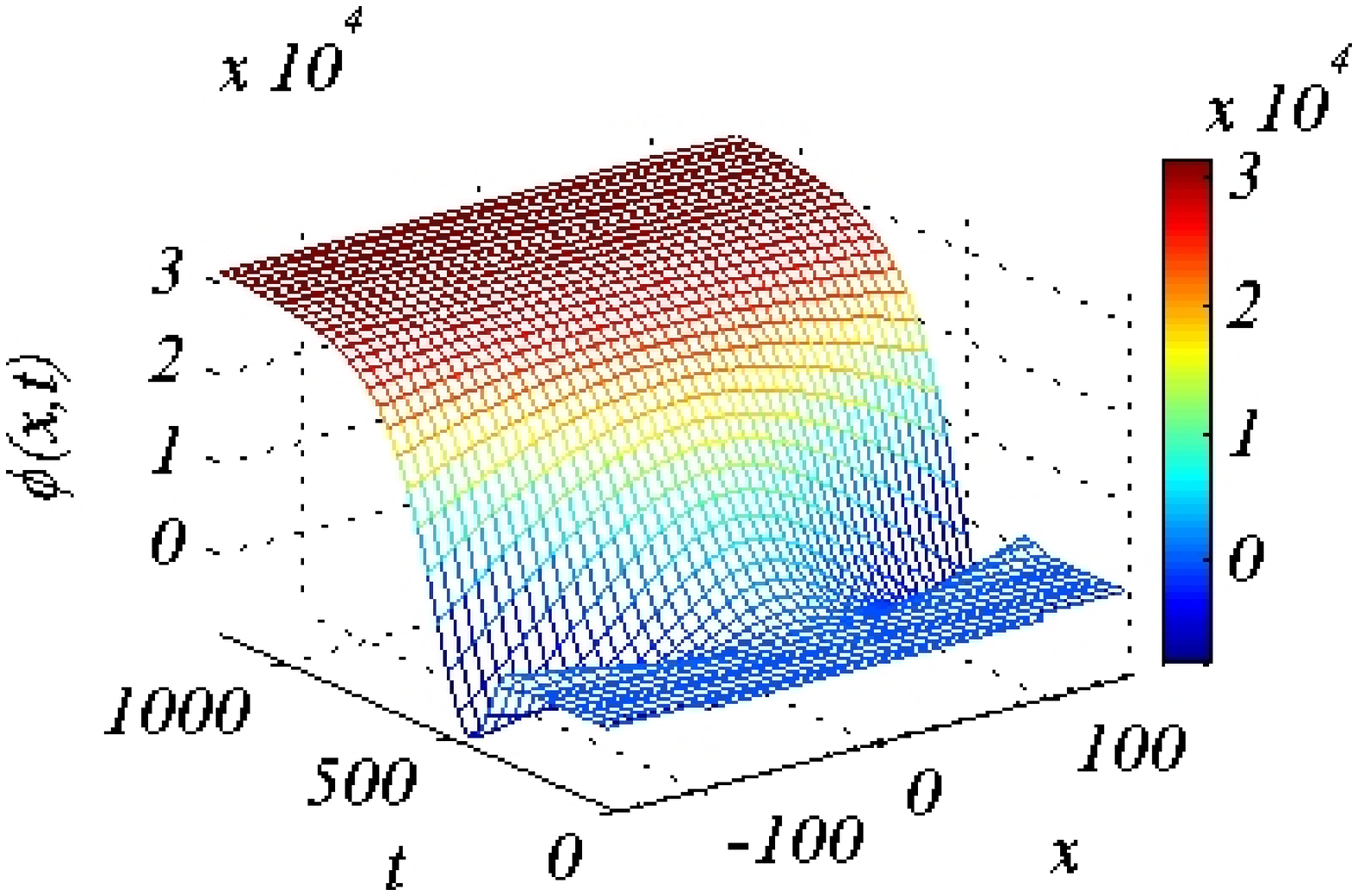} &
  \solidspace{5mm} &
  \includegraphics[height=40mm,width=60mm]{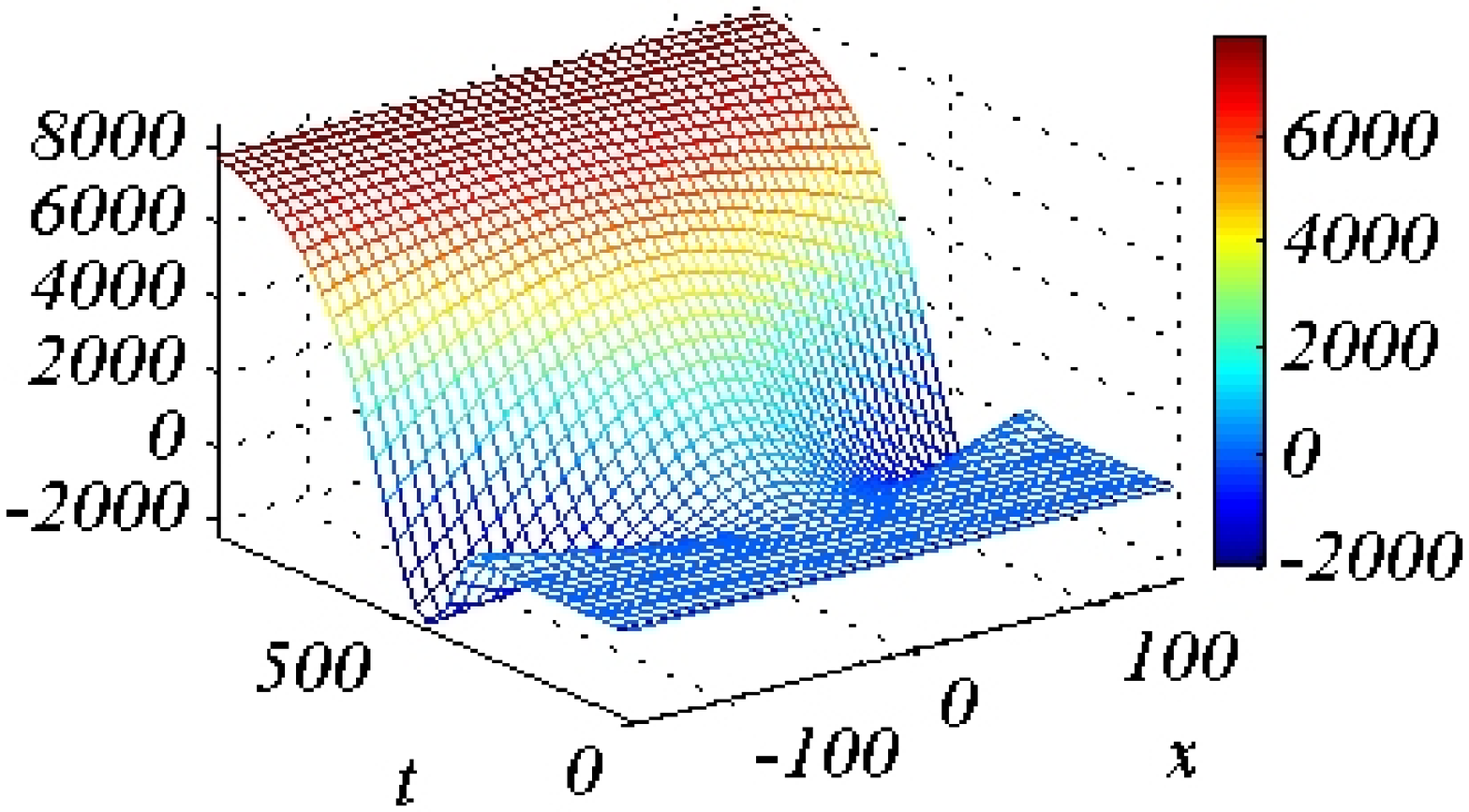}\\
  \solidspace{5mm}(c) $p=0.0765$ & & 
  \solidspace{5mm}(d) $\phi(x,0)=\delta(x)$\\
   & & \\
\end{tabular}
\end{center}
\caption{Examples of solutions \eq{gaussol2}, \eq{gaussol1} 
         for $R=146.7126$, which corresponds to $L=40\pi$.}
         \label{gaussol}
\end{figure}

Solution \eq{convsol}, \eq{funG1} for $\phi(x,0)=e^{-px^{2}}$, 
$p>0$ can be represented in a closed form as well. Routine 
integration yields
\bequ
\phi(x,t)=\frac{\pi}{\sqrt{pa}}e^{\frac{t}{R}+\frac{b^{2}-4\pi^{2}x^{2}}{4a}}
\left\{\cos\frac{\pi bx}{a}
+\Re\left[e^{\frac{i\pi bx}{a}}\erf\left(\frac{b+i2\pi x}{2\sqrt{a}}\right)\right]\right\},
\eequ{gaussol2}
where 
\bequ
a=a(t)=2\pi^{2}R\left(e^{2t/R}-1\right)+\pi^{2}e^{2t/R}/p,\qquad
b=b(t)=\pi\gamma R\left(e^{t/R}-1\right).
\eequ{gaussol1}
The result is illustrated in Fig. \ref{gaussol}, where case 
(c) corresponds to the maximum growing perturbation of type 
$\phi(x,0)=e^{-px^{2}}$ and initial condition 
$\phi(x,0)=\delta(x)$ was used in (d). The solution formula 
in the latter case is given by \eq{gaussol2} with $p$ formally 
replaced by $\pi$ and it also should be used with 
$a=a(t)=2\pi^{2}R\left(e^{2t/R}-1\right)$ and $b$ exactly the 
same as in \eq{gaussol1}.

The steady coalescent pole solutions to the Sivashinsky equation 
correspond to the flame fronts propagating steadily with the velocity
exceeding $u_{b}=1$ by $V_{L}$, see e.g. 
\cite{Rahibe-Aubry-Sivashinsky98}. Addition of the perturbation 
$\phi(x,t)$ results in a change in the velocity of propagation by the 
value of the space average of $\partial_{t}\phi$, which we denote
$\xavr{\phi_{t}}$. The correction provided by the $\phi(x,t)$ 
is only valid in a small vicinity of the crest 
of $\Phi_{L}(x,t)$ . In sequel, the correction of the speed $\xavr{\phi_{t}}$ 
is only valid for a small region $-\varepsilon\le x\le\varepsilon$ 
of the flame front in a vicinity of the crest of $\Phi_{L}(x,t)$. 
Hence, for our simplified linear model we define the increase 
of the flame propagation speed as follows: 
\bequ
\xavr{\phi_{t}}=\frac{1}{2\varepsilon} 
\int\limits_{-\varepsilon}^{\varepsilon}\partial_{t}\phi dx
\approx\left.\partial_{t}\phi\right|_{x=0}. 
\eequ{speeddef} 

\begin{figure}[ht] 
\centerline{\includegraphics[height=40mm,width=75mm]{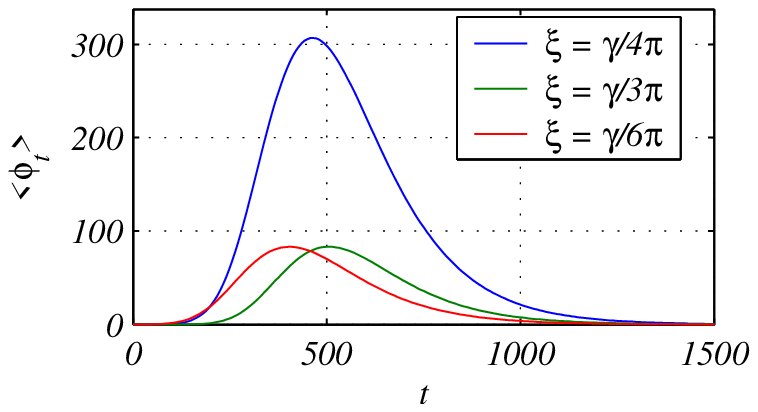}}
\caption{Averaged increase of the local flame propagation speed 
$\xavr{\phi_{t}}$ for solutions \eq{cosphisol}, $\varphi=0$.}
\label{cosspeedgr}
\end{figure}

For the single harmonics solution \eq{cosphisol} the expression 
for $\xavr{\phi_{t}}$ is obvious and is illustrated in Fig. \ref{cosspeedgr}. 
These graphs demonstrate high sensitivity of $\xavr{\phi_{t}}$ to 
the wavelength of the perturbation. The phase, or location of the 
perturbation, is important as well. 
 
\section{The effect of noise}
According to the results of Section \ref{WorstPert}, the forcing 
in the Sivashinsky equation can be decomposed into the most
amplifiable nonmodal component and the orthogonal complement.
The latter can be neglected reducing spatio-temporal 
stochastic noise to the appearance of a sequence of the most
growing perturbations $\psi_{\alpha_{m}}(x,t^{*})$, 
$1\le\alpha_{m}\le\alpha^{*}=\alpha^{*}(f_{0})$ at a set of 
time instances $t_{m}$, $m=0,1,2,\ldots$:
\bequ
f(x,t)\approx f_{0}\sum\limits_{m=0}^{\infty}
\psi_{\alpha_{m}}(x,t^{*})\delta(t-t_{m}),\qquad
1\le\alpha_{m}\le\alpha^{*}=\alpha^{*}(f_{0}).
\eequ{NoiseModel0}
Thus, the amplitude of noise $f_{0}$, alongside with the
averages and the standard deviations of $t_{m+1}-t_{m}$ 
and $\alpha_{m}$, $m=0,1,2,\ldots$ are the only essential 
parameters of such a representation of noise.

The impulse-like noise \eq{NoiseModel0} is used here for the 
sake of simplicity. Some arguments towards its validity were 
suggested in \cite{Joulin88}. More sophisticated and 
physically realistic models of temporal noise characteristics 
can be used with \eq{SivaEq} as well.

If $f_{0}\ll\sigma_{1}^{-1}(t^{*})$ then noise is not able to 
affect the flame at all and can be completely neglected. This 
case can be referred to as the noiseless regime. On the other 
hand, if $f_{0}$ is comparable with the amplitude $a$ of the 
background solution $\Phi_{L}(x,t)$, then almost all 
components of noise will be able to disturb the flame and 
the $f(x,t)$ in \eq{SivaEq} should be treated as a genuine 
spatio-temporal stochastic function. This is the regime of the 
saturated noise. 

Eventually, there is an important transitional regime when the noise 
amplitude $f_{0}$ is at least of order of $\sigma_{1}^{-1}(t^{*})$, 
but still much smaller than $a$. In this case only the disturbances 
with a significant component in the subspace spanned by the linear 
combinations of $\psi_{\alpha}(x,t^{*})$, $1\le\alpha\le\alpha^{*}$ 
have a potential to affect the solution. All other disturbances can 
be neglected and the force $f(x,t)$ in the Sivashinsky equation 
\eq{SivaEq} can be approximated by \eq{NoiseModel0} with a finite
value of $\alpha^{*}$. We would like to stress that though such 
representation of noise is correct for noise of any amplitude, 
apparently it is only efficient if 
$f_{0}<\sigma_{\alpha^{*}}^{-1}(t^{*})\ll a$, where $\alpha^{*}$ 
is small enough.

\subsection{Noise in the linear model}
A random point-wise set of perturbations uniformly distributed 
in time and in the Fourier space is a suitable model for both 
the computational round-off errors and a variety of perturbations 
of physical origins. We are adopting such a model in our analysis 
in the following form 
\bequ
f(x,t)=\sum\limits_{m=1}^{M(t)}a_{m}\cos(2\pi\xi_{m}x+\varphi_{m})
\delta(t-t_{m}),
\eequ{noise}
where $a_{m}$, $t_{m}$, $\xi_{m}$, and $\varphi_{m}$ are 
non-correlated random sequences. It is assumed that 
$t_{1}\le t_{2}\le\cdots\le t_{m}\le\cdots\le t_{M(t)}\le t$,
$0\le\varphi_{m}\le 2\pi$, and $\xi_{m}\ge 0$, $m=1,2,\ldots,M(t)$.
Availability of the exact solution \eq{gaussol2} makes it also 
possible to study an alternative noise model based on elementary 
perturbations $a_{m}e^{-p_{m}(x-x_{m})^{2}}$, which are local in 
physical space.

\begin{figure}[ht]
\begin{centering}\hfill
\begin{tabular}{p{65mm}p{65mm}}
\includegraphics[height=40mm,width=65mm]{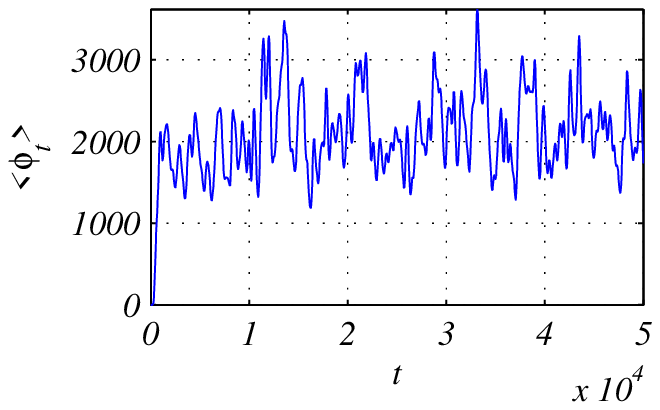} &
\includegraphics[height=40mm,width=65mm]{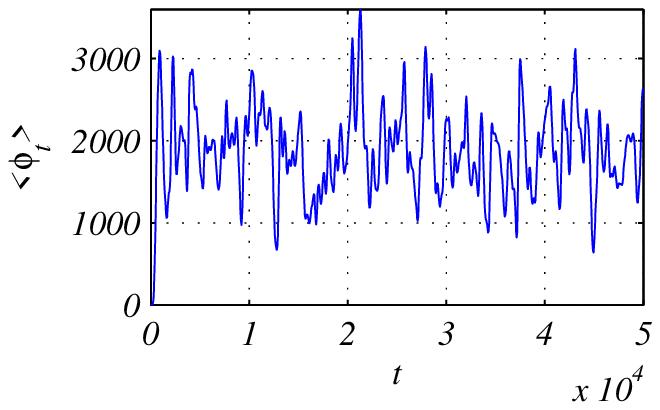}\\
\solidspace{5mm}(a) $F=1$,\ \ $a_{m},\xi_{m}\in[0,1]$ & 
\solidspace{5mm}(b) $F=1/33$, $a_{m}\in[0,1]$,\\ 
& \solidspace{12mm}$\xi_{m}\equiv \gamma/(4\pi)$\\
\end{tabular}\hfill
\end{centering}
\caption{The effect of noise \eq{noise} on $\xavr{\phi_{t}}$ 
         for $L=40\pi$.}\label{noise1i5}
\end{figure}

Using \eq{convsol}, \eq{funG1} for the zero initial condition, 
the exact solution to \eq{Joulin}, \eq{noise} can be written as 
$$
\phi(x,t)=\sum\limits_{m=1}^{M(t)}a_{m}
e^{-2\pi^{2}R\left[1-e^{-2(t-t_{m})/R}\right]\xi_{m}^{2}
+\pi\gamma R\left[1-e^{-(t-t_{m})/R}\right]\xi_{m}}
$$
\bequ
\times\cos\left[2\pi\xi_{m}e^{- (t-t_{m})/R}x+\varphi_{m}\right].
\eequ{noisesolfin} 
The expression for $\xavr{\phi_{t}}$ is obvious, see 
\eq{speeddef}, and is illustrated in Fig. \ref{noise1i5}.
Here we generated random sequences of the time instances 
$t_{m}$ with a given frequency $F=M(T)/T$ on a time 
interval $t\in[0,T]$. Values of the wave number $\xi_{m}$ 
and of the amplitude $a_{m}$ were also randomly generated 
and uniformly distributed within certain ranges. According 
to the formula for $\xavr{\phi_{t}}$, the effect of the 
phase shift $\varphi_{m}$ just duplicates the $a_{m}$. 
Therefore, its value was fixed as $\varphi_{m}\equiv 0$. 

If values of $a_{m}$ are uniformly distributed in $[-1,1]$, 
then the time average of $\xavr{\phi_{t}}$ is obviously zero, 
because of the linearity of the problem. In the Sivashinsky 
equation this effect is compensated by the nonlinearity. The 
cusps generated by the perturbations of opposite signs move 
into opposite directions along the flame surface, see Section 
\ref{WorstPert}, though they both contribute into the speed 
positively. This effect of the nonlinearity can be mimicked 
by restricting the range of possible values of the amplitudes, 
e.g. $a_{m}\in[0,1]$, as this can be seen in Fig. \ref{noise1i5}.

Figure \ref{noise1i5}(b) shows that the increase of 
$\xavr{\phi_{t}}$ seen in Fig. \ref{noise1i5}(a) can be 
matched by using only the largest growing perturbations with 
much smaller frequency, which is quite expected in virtue of 
the linearity of the problem. The amplitude of fluctuations 
in Fig. \ref{noise1i5}(a) is noticeably less than in Fig. 
\ref{noise1i5}(b). This is attributed to the smoothing effect 
of the less growing perturbations. 

\begin{figure}[ht]
\centerline{\includegraphics[height=40mm,width=70mm]{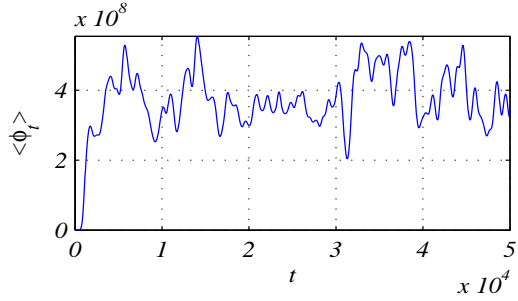}}
\caption{The effect of noise \eq{noise} on $\xavr{\phi_{t}}$ 
         for $L=80\pi$, $F=1$, $a_{m}\in[0,1]$, and 
         $\xi_{m}\in[0,1]$.}\label{noise2}
\end{figure}

Because of the linearity of the problem in question, the 
effect of $F=M(T)/T$ and $L$ on $\xavr{\phi_{t}}$ is 
straightforward. In particular, the value of $\xavr{\phi_{t}}$ 
raises up to about $4\times 10^{8}$ for $L=80\pi$ and other 
parameters the same as in Fig. \ref{noise1i5}(a). It should 
be noticed however that because of the limitation $a_{m}\ge 0$ 
the quantity $\xavr{\phi_{t}}$ does no longer represent the 
increase of propagation speed of the flame, but is just a 
measure of the rate of transient amplification of 
perturbations. Figure \ref{noise2} illustrates the point, 
see also \cite{Karlin02a} and \cite{Karlin-Maz'ya-Schmidt03}.

Direct studies of the effect of noise in the Sivashinsky 
equation necessitate use of numerical simulations. However, 
because of the intrinsic discontinuity of noise, such DNS 
are hampered with very low accuracy of approximations 
questioning the validity of numerical solutions. In this 
work we used explicit solutions \eq{noisesolfin} in order 
to validate DNS of \eq{Joulin} and, in sequel, of \eq{SivaEq}. 
The DNS of \eq{Joulin}, \eq{noise} was carried out using a 
spectral method. The delta function $\delta(t-t_{m})$ was 
approximated by $(\pi\Delta t)^{-1/2}e^{-(t-t_{m})^{2}/\Delta t}$ 
with a small enough $\Delta t\ll 1/F$. The calculations 
have shown that discrepancies between \eq{noisesolfin} and 
its numerical counterparts obtained with the same sets of 
$t_{m}$, $a_{m}$, and $\xi_{m}$ might be noticeable in a 
neighbourhood of the time instances $t\approx t_{m}$. 
Although, the averaged characteristics like $\xavr{\phi_{t}}$ 
were quite accurate. So, this linear model validates the 
DNS of the forced Sivashinsky equation at least in 
relation to the averaged flame propagation speed.

\subsection{The Sivashinsky equation}
We carried out a series of computations of \eq{SivaEq}, 
\eq{NoiseModel0} with $\Phi_{L}(x,t)$ as initial condition 
and with a variety of parameters of the noise term. Up 
to twelve basis functions $\psi_{\alpha}(x,10^{3})$, where 
$\alpha$ was uniformly distributed in the interval 
$1\le\alpha\le\alpha^{*}\le 12$, were used. The sign of $f_{0}$ in 
\eq{NoiseModel0} was either plus or minus for every $m$ 
with the equal probability $1/2$. The delta function 
$\delta(t-t_{m})$ was approximated by 
$(\pi\Delta t)^{-1/2}e^{-(t-t_{m})^{2}/\Delta t}$ with a 
small enough value of $\Delta t$.

\begin{figure}[ht]\centering
  \includegraphics[height=40mm,width=75mm]{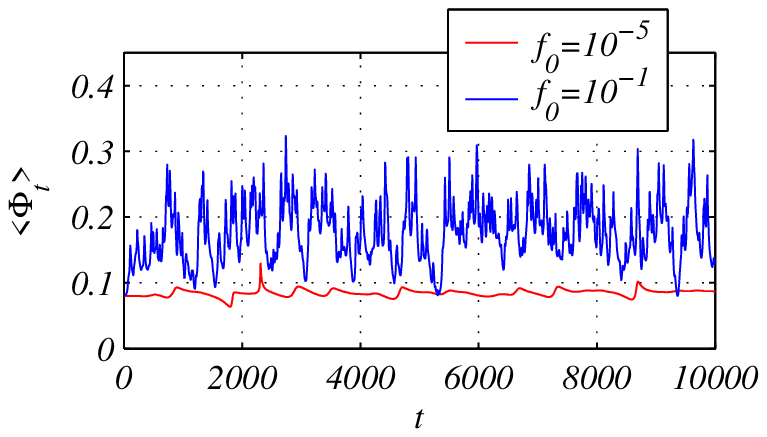}
  \caption{The effect of the amplitude of noise \eq{NoiseModel0} 
           on $\xavr{\Phi_{t}}$ for $L=40\pi$. Here $F=1/15$ and 
           $\alpha^{*}=12$.}
  \label{siva_noise1}
\end{figure}

\begin{figure}[ht]\centering
  \includegraphics[height=120mm,width=120mm]{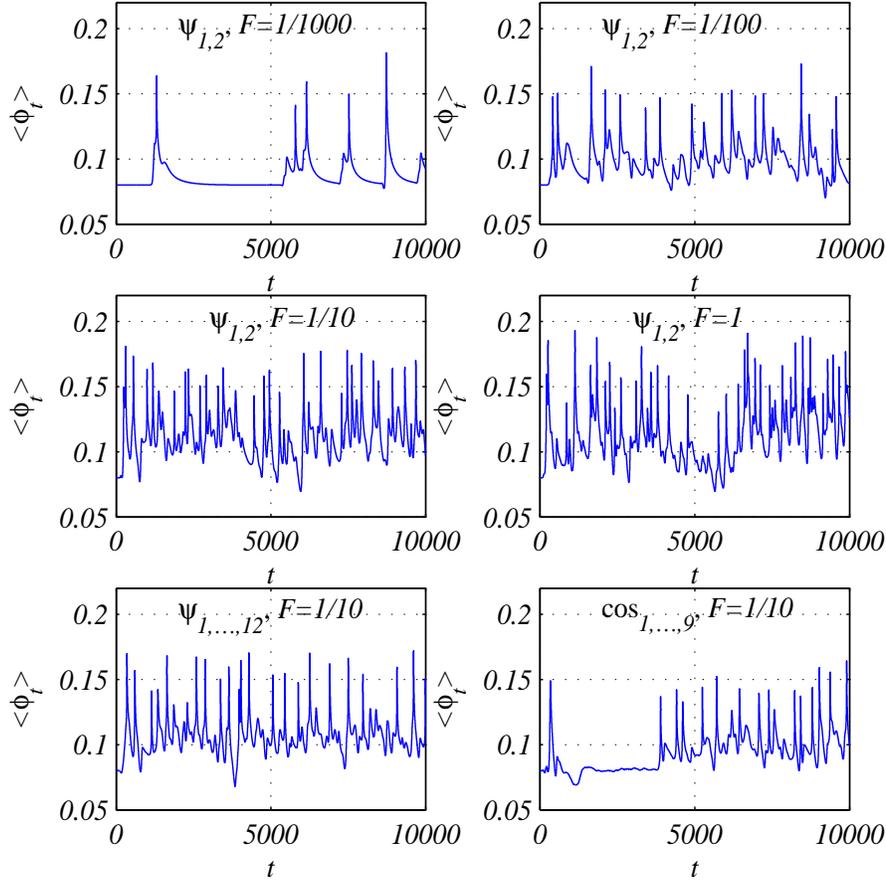}
  \caption{Examples of the effect of the composition of noise on 
           $\xavr{\Phi_{t}}$ for $f_{0}=10^{-3}$ and $L=40\pi$.}
  \label{siva_noise1b}
\end{figure}

The effect of the amplitude of noise on the flame speed is 
illustrated in Fig. \ref{siva_noise1}. Use of only two basis 
functions $\psi_{1,2}(x,10^{3})$ gives almost the same 
result. Similar to the linear model \eq{Joulin} the only 
noticeable difference was in slightly larger fluctuations 
of $\xavr{\Phi_{t}}$. Examples of the effect of composition 
of noise on $\xavr{\Phi_{t}}$ are given in Fig. \ref{siva_noise1b}
too. 

It was mentioned in the previous section that the wave number 
of the largest Fourier component of $\psi_{1,2}(x,10^{3})$ is 
exactly the same as the wave number $\xi_{0}=\gamma/(4\pi)$ of 
the largest growing single harmonics solution to \eq{Joulin}. 
We tried to exploit this observation and simplified 
\eq{NoiseModel0} even further, replacing 
$\psi_{\alpha(m)}(x,t^{*})$ by $\cos(\gamma x/4)$, which 
corresponds to $\xi_{0}$, i.e. 
\bequ
f(x,t)\approx f_{0}\cos\frac{\gamma x}{4}
\sum\limits_{m=0}^{\infty}\delta(t-t_{m}).
\eequ{NoiseModel1}
This kind of forcing is able to speed up the flame, but the 
difference between the computational results obtained with 
\eq{NoiseModel0} and \eq{NoiseModel1} is noticeable. It does 
not disappear even if eight nearest sidebands are added to 
\eq{NoiseModel1}, see Fig. \ref{siva_noise1b}.

\begin{figure}[ht]
\begin{centering}\hfill
\begin{tabular}{p{75mm}p{55mm}}
\includegraphics[height=45mm,width=75mm]{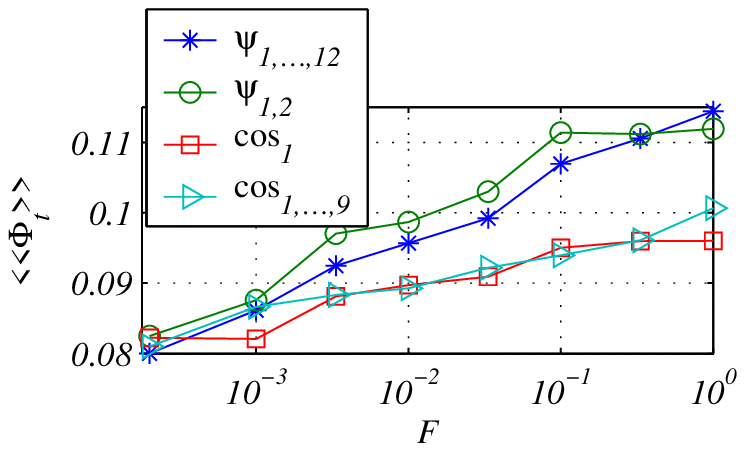} &
\includegraphics[height=40mm,width=55mm]{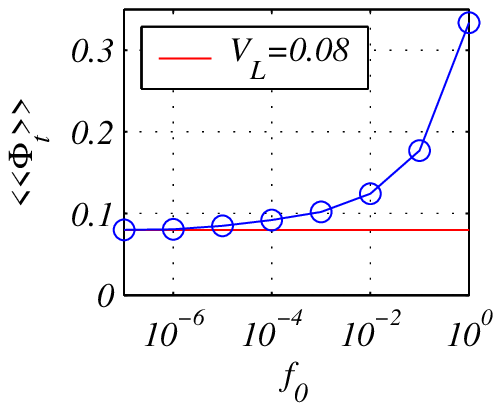}\\
\solidspace{5mm}(a) $f_{0}=10^{-3}$ & 
\solidspace{5mm}(b) $F=1/15$, $\alpha^{*}=12$,\\ 
\end{tabular}\hfill
\end{centering}
\caption{The effect of the composition (a) and amplitude (b) 
of noise on the spatio-temporally averaged flame propagation 
speed $\xtavr{\Phi_{t}}$. Here $L=40\pi$ and the temporal 
averaging was over the interval 
$t\in[200,10^{4}]$.}\label{siva_noise2}
\end{figure}

The time averages of $\xavr{\Phi_{t}}$, denoted here as 
\bequ 
\xtavr{\Phi_{t}} = \frac{1}{t_{end}-t_{0}}\int\limits_{t_{0}}^{t_{end}}
\xavr{\Phi_{t}}dt, 
\eequ{FlameSpeed1} 
are depicted in Fig. \ref{siva_noise2} versus $F$ (left) 
and $f_{0}$ (right).
Discrepancies in $\xtavr{\Phi_{t}}$ for different $\alpha^{*}$ 
did not exceed the variations caused by the different randomly 
chosen sequences of $t_{m}$. Although, effect of using 
\eq{NoiseModel1} instead of \eq{NoiseModel0} is appreciable.

The correlation between the flame propagation speed and the 
noise amplitude is obvious. Note that the $f_{0}$ in the right 
most point in the graph is still about $20$ times less than 
the amplitude of the variation of the background solution 
$\Phi_{L}(x,t)$. 

In accordance with the idea developed in this paper, the value of 
$\xtavr{\Phi_{t}}$ is determined by the product $\sigma_{1}f_{0}$. 
It was shown in \cite{Karlin02a} that $\sigma_{1}\propto e^{O(L)}$, 
resulting in \mbox{$\xtavr{\Phi_{t}}=\xtavr{\Phi_{t}}(e^{O(L)}f_{0})$}. 
Thus, the data shown in Fig. \ref{siva_noise2} are at least in a 
qualitative agreement with the dependence of $\xtavr{\Phi_{t}}$ on 
$L$, which was obtained in \cite{Karlin02a} for a fixed noise 
amplitude $f_{0}\approx 10^{-16}$ associated with the computational 
round-off errors. 

\begin{figure}[ht]
\begin{centering}\hfill
\begin{tabular}{p{65mm}p{65mm}}
  \includegraphics[height=40mm,width=65mm]{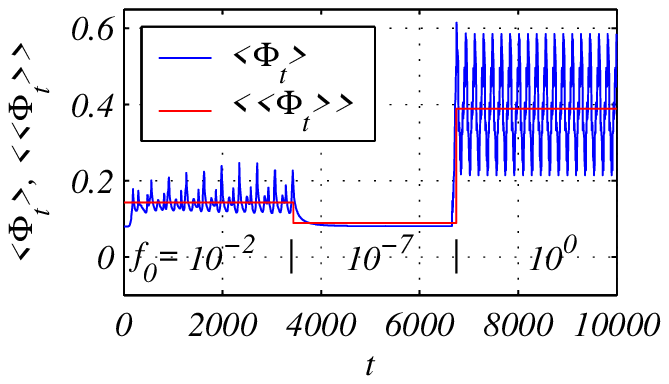} &
  \includegraphics[height=40mm,width=65mm]{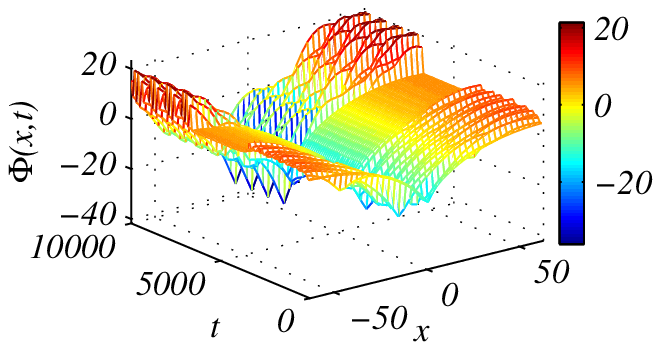}\\
\end{tabular}\hfill
\end{centering}
\caption{An example of controlling the flame speed with the amplitude 
of the perturbations $f_{0}$. Here $F=1/15$, $\alpha^{*}=12$,
and $L=40\pi$.}\label{siva_cont}
\end{figure}

Eventually, in Fig. \ref{siva_cont} we presented the results of an 
attempt to control the flame propagation speed using our special 
perturbations $\psi_{\alpha}(x,t^{*})$ of properly selected 
amplitudes. Graphs of $\xavr{\Phi_{t}}$ and $\xtavr{\Phi_{t}}$ are 
shown in the left and numerical solution $\Phi(x,t)$ corresponding 
to this controlling experiment is illustrated in the right. The 
fluctuations of the obtained flame propagation speed are large indeed,
but at least, they appear in quite a regular pattern.

In this paper noise or forcing in \eq{SivaEq} represents 
the turbulence of the upstream velocity field, which is 
difficult to manage in practice. The controlling function is 
more effectively achieved by acoustic signals, see e.g. 
\cite{Clanet-Searby98}. Acoustics was neglected in the 
evaluation of the Sivashinsky equation and there is no easy 
and straightforward way to incorporate it back into the model. 
However, because of a strong coupling between the velocity 
and pressure fields, effects of acoustic signals similar to 
those presented here can be expected as well.

\section{Conclusions}
Based on our analysis of the steadily propagating cellular 
flames governed by the Sivashinsky equation we may conclude 
that there are perturbations of very small amplitude, which 
can essentially affect the flame front dynamics. The subspace 
formed by these special perturbations is of a very small 
dimension and its basis can be used for an efficient 
representation of the upstream velocity turbulence. These are 
the very perturbations which cause the increase of the flame 
propagation speed in numerical experiments. Hence, 
theoretically, they can be used to model certain regimes of 
flame-turbulence interaction and to control the flame 
propagation speed on purpose.

\section*{Acknowledgements}
The research presented in this paper was supported by the EPSRC 
grant\linebreak GR/R66692.

\bibliography{pap5a}

\end{document}